\documentclass[11pt]{article} 

\usepackage{contigrefs}
\usepackage[utf8]{inputenc} 

\usepackage{geometry} 
\usepackage{graphicx} 

\usepackage{booktabs} 
\usepackage{array} 
\usepackage{paralist} 
\usepackage{verbatim} 
\usepackage{subfig} 
\usepackage{amsmath} 
\usepackage{amssymb} 
\usepackage{mathtools} 
\DeclarePairedDelimiter{\ceil}{\lceil}{\rceil}

\usepackage{epstopdf} 
\usepackage{color}

\usepackage{algorithm} 
\usepackage[noend]{algpseudocode}
\def\BState{\State\hskip-\ALG@thistlm}

\usepackage{fancyhdr} 
\usepackage{sectsty}
\allsectionsfont{\sffamily\mdseries\upshape} 

\usepackage[nottoc,notlof,notlot]{tocbibind} 
\usepackage[titles,subfigure]{tocloft} 

\title{Influence Maximization for Fixed Heterogeneous Thresholds}
\author{P. D. Karampourniotis$^{1,2,3}\footnote{E-mail: karamp.pan@gmail.com}$,  B. K. Szymanski$^{3,4,5}$, G. Korniss$^{2,3}$}

\begin{document}
\maketitle

\begin{flushleft}
$^{\bf{1}}$ IBM Research, Cambridge, MA, USA \\
$^{\bf{2}}$ Department of Physics, Applied Physics, and Astronomy, Rensselaer Polytechnic Institute,
110 8$^{th}$ Street, Troy, NY, 12180-3590 USA \\
$^{\bf{3}}$ Social Cognitive Networks Academic Research Center,
Rensselaer Polytechnic Institute, 110 8$^{th}$ Street, Troy, NY, 12180-3590 USA \\
$^{\bf{4}}$ Department of Computer Science,
Rensselaer Polytechnic Institute, 110 8$^{th}$ Street, Troy, NY, 12180-3590 USA \\
$^{\bf{5}}$ Faculty of Computer Science and Management, Wroclaw University of Science and Technology, Poland
\\
\end{flushleft}

\section*{Abstract}
Influence Maximization is a NP-hard problem of selecting the optimal set of influencers in a network. Here, we propose two new approaches to influence maximization based on two very different metrics. The first metric, termed Balanced Index (BI), is fast to compute and assigns top values to two kinds of nodes: those with high resistance to adoption, and those with large out-degree. This is done by linearly combining three properties of a node: its degree, susceptibility to new opinions, and the impact its activation will have on its neighborhood. Controlling the weights between those three terms has a huge impact on performance. The second metric, termed Group Performance Index (GPI), measures performance of each node as an initiator when it is a part of randomly selected initiator set. In each such selection, the score assigned to each teammate is inversely proportional to the number of initiators causing the desired spread. These two metrics are applicable to various cascade models; here we test them on the Linear Threshold Model with fixed and known thresholds. Furthermore, we study the impact of network degree assortativity and threshold distribution on the cascade size for metrics including ours. The results demonstrate our two metrics deliver strong performance for influence maximization.


\section*{Introduction}

Cascading processes emerge naturally through the interactions of nodes 
in different states in natural and human-made networks. Microscopic 
processes can potentially have large macroscopic impact on the networks.
In the case of human-made networks, their ever increasing size and 
interconnectedness exponentially increases the uncanny impact of cascade processes. 
For instance, in financial or power grid networks, small initial perturbations 
or failures can result in cascades in the network causing tremendous disasters of 
global impact~\cite{Crucitti_PhysA,Kinney_EPJB,Hines}. In social networks, contact processes, namely social 
influence (or contagion), enable the spread of new behaviors, opinions 
and products, driving politics, social movements and norms, and viral marketing. 

The key challenge in predicting such cascading processes is the identification of nodes whose change of state can potentially 
affect large portions of the networks. It is a computationally 
hard problem, and as such, multiple heuristics, theoretical analyses and algorithms have 
been introduced to solve it~\cite{Zhou_PRE,Zhou_arXiv,Braunstein_PNAS}. 
Some are designed to address the specific nature of 
the cascade process, while others are based on more general algorithmic approaches 
or network based centrality measures. Such algorithms can be used to minimize disasters 
by, for example, re-enforcing weak nodes in power-grid nodes~\cite{Motter_PRE,Wang_Safety,Asztalos_PLOS}, 
or placing sensors to detect the contamination in water pipe network~\cite{Leskovec_CELF}. 
Likewise, arresting the spread of infectious disease requires a global awareness of its existance~\cite{Kitsak_NatPhys2010,Pastor_RevModPhys,Altarelli_PhysRevX,Malliaros}. 
Understanding cascades is also important for optimizing viral marketing~\cite{Domingos_2001,Domingos_2002,Barbieri}.
Yet, it is challenging to find the set of 
initiators (also called seeds) which, when put into a new state (opinion/idea/product),
will maximize the spread of this state~\cite{Kempe_2003,Chen_2010,Goyal_CELF++,Goyal_SIMPATH,Chen_2012,Jankowski_SciRep,Jiang_AAAI,Lu_PhysRep,Shakarian_SNAM,Morone_Nature,Makse_CITM}.

Here, we study the problem of influence maximization for a classical opinion contagion model, namely the Linear Threshold Model (LTM)~\cite{Granovetter,Schelling}. 
Yet our methods can be used for any percolation based model.
The LTM is designed to capture the peer pressure dynamics that lead an 
individual to accept a new state being propagated.
It is a binary state model, where a node $i$ has either adopted a new product/state/opinion, $n_i=1$, or not, $n_i=0$.
In the LTM, each node in the network has a fractional threshold, 
an intrinsic parameter representing the node's resistance to peer pressure.
The spreading rule is that an inactive node ($n_i=0$), with in-degree $k^{{\rm in}}_{i}$ and threshold $\phi_i$,
adopts a new opinion only when the fraction of its neighbors $j \in \partial i$ holding 
the new opinion is higher than the node's threshold, that is $\sum_{j \in\partial i} n_j \ge\phi_ik^{{\rm in}}_{i}$. The process is deterministic 
and once a node adopts a new opinion it cannot return to its previous state. The integer number of active neighbors required for node $i$ to get active is given by its initial resistance $r_i=\ceil{\phi_ik^{{\rm in}}_{i}}$. Naturally, the maximum resistance of a node is $k^{in}$. At each step of the spread, the resistance is decreased by the number of the node's neighbors that adopted a new opinion.
A node gets activated through spread when its resistance drops to zero or below, $r_i\leq0$. Bootstrap percolation~\cite{Baxter_PRE2010,Baxter_PRE2011} is an alternative formulation of the LTM where the thresholds are not fractional, but integer (resistance).
Interestingly, LTM and bootstrap percolation conceptually share similarities with the integrate-and-fire-neuron model~\cite{Amini,Eckmanna,Kozma,Lymperopoulos},
with the main differences being that in the later, activation of a node is probabilistic rather than a fixed threshold value, and return to the initial (inactive) state $n_i = 0$ is allowed.
The size of cascades in the LTM is governed by the thresholds of the nodes~\cite{Karamp_PLOSONE,Watts_2007},
the size of the initiator set~\cite{Singh}, the selection of initiators~\cite{Makse_CITM}, and of course the structure of the underlying network~\cite{Gleeson_2008,Ikeda_2010,Lee_2014,Nematzadeh_2014,Curato_PRE}.

Examples of an LTM type spread mechanisms and of the heterogeneity of the thresholds are provided through a number of controlled experiments~\cite{Latane_JPS1996,Centola_2010,Monsted_arXiv} and by empirical data analysis~\cite{Romero,Karsai_SciRep2016,Fink_SNAM,Fink_AAAI,State_2015}. Unicomb et al. \cite{Unicomb_2018} studied the threshold model in weighted networks and found  that the time of cascade emergence depends non-monotonically on weight heterogeneities. Watts and Dodds~\cite{Watts_2007} showed through simulations of various types of spread mechanisms that the cascade size is governed not by superspreaders, but by a small critical set of nodes with low resistance to influence.
Karampourniotis et al. showed that the threshold distribution is important for the overall dynamics~\cite{Karamp_PLOSONE}. In particular, with an increasing standard variation $\sigma$ of thresholds (while holding the average of the thresholds fixed) a smaller initiator fraction is required for global cascades. Furthermore, they showed that in the vicinity of $\sigma\approx 0$ a tipping point appears as the required fraction of randomly selected initiators gradually increases. Yet, with gradually increasing variance $\sigma$, eventually the tipping point is replaced by smooth transition.
In addition, Watts and Dodds~\cite{Watts_2007} showed that a critical fraction of nodes with high susceptibility contributes to social influence much more than initiators with high network centrality.

Hence, the knowledge of the thresholds or their distribution is critical for the influence maximization algorithms. 
In the case of zero information about the thresholds or their distribution, a good assumption to make 
is that the threshold distribution is uniform. This is a very interesting case for which the spread function is submodular,
that is, it follows a diminishing returns property~\cite{Kempe_2003}, which can be used for maximizing the influence~\cite{Leskovec_CELF,Kempe_2003,Chen_2010,Goyal_CELF++,Goyal_SIMPATH,Jiang_AAAI,Tang_ACM,Liu_IEEE}. In general,
submodularity does not hold when the thresholds are known and fixed or for any threshold distribution but the uniform. 
For the cases of fixed known thresholds, the influence of any seed set can be computed in polynomial time~\cite{Lu_Optim}.
Further, in the special case of a unique threshold for all nodes, a first-order transition appears~\cite{Singh,Watts_2002,Gleeson_2007}. Then a powerful algorithm, namely CI-TM with complexity $\mathcal{O}\left(\langle k\rangle N \log N\right)$, provides great performance~\cite{Makse_CITM}.
In Ref.~\cite{Altarelli_JStatMech} the authors express the influence maximization problem as a constraint-satisfaction problem and use believe propagation to solve it. Yet, for the case of identical thresholds it does not perform as 
well as the CI-TM with which it was compared in~\cite{Makse_CITM}. Other analytical based metrics show the importance of network structure, but only for a small number of initiators \cite{Lim_2015}. 
Furthermore, in~\cite{Michalski_ASONAM} the authors propose to use of an evolutionary algorithm 
implemented with general-purpose computing on graphics processing units (GPGPU) to tackle the challenge of combinatorics, at the additional cost 
of higher time and memory complexity. The authors show that their approach clearly outperforms the greedy algorithm 
for known thresholds both in cascade size and time, but it is currently limited to graphs of size in the order of $N=10^3$.

Here, we study the efficiency of known selection strategies for fixed (and known) heterogeneous thresholds generated from different threshold distributions, and a range of assortativity values. We introduce two new and very different selection strategies for the LTM with fixed thresholds, and compare them in terms of their performance with a number of other strategies, including the CI-TM, and greedy. Since we focus on fixed (and known) thresholds we do not include the performance of various network centrality measures like the Page Rank and $k$-core~\cite{Singh}, which do not take into account the provided threshold information and thus are outperformed by the strategies that do.

\section*{Selection Strategies}
We use many fast heuristics, which take advantage of the
knowledge of thresholds. Since the thresholds are fixed and known, the cascade size 
caused by an initiator is deterministic. Hence, we sequentially introduce initiators {\it on 
the inactive subgraph} of the original network. First, the node with the highest dynamic fractional 
threshold ($thres$) is a reasonable choice. Likewise, a natural selection is the node with the highest dynamic out-degree $k^{\rm{out}}_{i}$ (i.e., counting only edges to currently inactive nodes) at each step ($deg$).
Another possible heuristic for the LTM is the selection of the node with the highest resistance ($res$) 
at each step. Resistance $r_i$ is the current (dynamic) integer threshold of node $i$, defined as the number of 
active neighbors required for the node to get activated. Accordingly, when a node is activated by a cascade or 
by being selected as an initiator, its resistance turns zero, and so a fully activated network
has total resistance of zero.

The selection of any inactive node $i$ as an initiator lowers the resistance of all its inactive neighbor nodes by one, for a total decrease of $k^{{\rm out}}_{i}$. Moreover, if any neighbor of $i$ had resistance equal to one, it will also get activated, further reducing the resistance of other nodes. In addition, if node $i$ is the initiator, then its resistance is reduced to zero, $r_i=0$. Therefore, the total reduction of the entire graph resistance is at least $r_i+k^{{\rm out}}_{i}$. To capture the resistance drop at the direct neighbors of node $i$, termed direct drop (DD), we introduce the heuristic strategy DD, with metric DD$_i=r_i+k^{{\rm out}}_{i}$. 
In addition, we introduce the heuristic method indirect drop of resistance (ID), by summing the DD metric and the drop of the resistance of the inactive subgraph caused by the neighbors of the node $i$ which are indirectly activated by the selection of $i$ as seed.
That is ID$_{i}=r_i+k^{\rm{out}}_{i}+\sum_{j\in \partial i |r_j=1}(k^{{\rm out}}_{j}-1)$. The total drop of the network's resistance caused by choosing $i$ as an initiator is at least as high as ID$_i$; it is more if the spread expands beyond the direct neighborhood (e.g the second or further neighborhoods).
The ID heuristic is equivalent to the CI-TM algorithm~\cite{Makse_CITM} for a sphere of influence $L=1$ with the addition of the resistance, $r_i$, of the selected node $i$ in the metric. For very large $L$, a node's $i$ CI-TM score is essentially (assuming a tree-like approximation of the network) equal to the drop of the network's total resistance if $i$ was the seed, minus node's $i$ resistance $r_i$. The metric of CI-TM is governed by the out-degree of the nodes surrounding the target node ignoring the challenge of activating nodes with high resistance and/or low in-degree.
For comparison to our methods, we apply the CI-TM algorithm itself (for $L=6$), and the greedy algorithm for fixed thresholds, where at each step the node which would cause the maximum cascade size is selected.

\subsection*{Balanced Index Strategy}
Constructing a selection strategy mainly based on the network structure or just on the resistance of the nodes, 
is not ideal, since useful information is not being utilized. On one hand, selecting nodes solely based on some network centrality metric, leads to many easily susceptible nodes being selected as initiators, nodes that could potentially be activated through spread. On the other hand, aiming at selecting high resistance nodes, does remove spread obstacles but does not guarantee they will be great influencers. 
The DD and ID strategies aim to address this weakness by using intuitive heuristics.
To quantify the interplay between the importance of low resistance and high centrality nodes, we introduce the Balanced Index (BI) selection strategy. For this strategy, we assign weights $(a,b,c)$ to each term of ID$_i$ to capture the importance of each feature, that is
\begin{equation}
\label{Eq1}
{\rm BI}_i=ar_i+bk^{{\rm out}}_{i}+c\sum_{j\in \partial i |r_j=1} \left(k^{{\rm out}}_{j}-1\right),
\end{equation}
where $a+b+c=1$ and $a, b, c\ge0$.
The optimal weights for influence maximization are determined by scanning within the ranges of weights in the ensemble of graphs and for various threshold distributions. In this case, the out-degree ({$deg$) strategy corresponds to $\left(a,b,c\right)=\left(0,1,0\right)$, $res$ to $\left(a,b,c\right)=\left(1,0,0\right)$, the CI-TM for $L=1$ to $\left(a,b,c\right)=\left(0,1/2,1/2\right)$, while the two heuristics we introduced, DD$_i$ and ID$_i$, correspond to weights $\left(a,b,c\right)=\left(1/2, 1/2, 0\right)$ and $\left(a,b,c\right)=\left(1/3, 1/3, 1/3\right)$ respectively. Interestingly, the weighted metric ${\rm BI}_i$ can be viewed as a measure (units) of resistance, however, in general, it does not correspond to the network's total resistance drop when $i$ is the seed.
As far as the time complexity of each method is concerned, the computation of a seed's induced spread takes $\mathcal{O}\left(\langle k\rangle N\right)$ time. Yet, (similar to~\cite{Makse_CITM}) when computing the spreading process, we can place a stop condition on the algorithm $L$ levels away from the seed node, reducing the complexity by $\mathcal{O}\left(N\right)$. In addition, using a heap structure, re-ordering the highest BI nodes takes $\mathcal{O}\left(\log N\right)$.

\subsection*{Group Performance Index algorithm} 
All of the above strategies are essentially local in nature, since they aim to maximize the number of activated nodes or to reduce the total resistance of the system caused by one initiator at a time. They lack in their metrics the impact of the combination of initiators on influence maximization, which by default limits their performance. Algorithms that use combinations of nodes in their metrics can lead to high-quality approximate global solutions. However, look-ahead algorithms suffer from the potentially prohibitive computational costs.
For instance, to measure the total impact of a subset of $g$ nodes selected from the total of $N$ nodes, a deterministic greedy algorithm would require ${N}\choose{g}$ selections of possible initiator sets.
The algorithm would choose iteratively at each step $\left(t=1,2\ldots ,g\right)$ the highest impact node from the set of the inactive ones, therefore its complexity would be $\mathcal{O}\left(\langle k\rangle N^g\right)$, feasible only for very small $g$ and moderate $N$. In addition, to compute the cascade size for each possible initiator set would require $\mathcal{O}\left(\langle k\rangle N^2\right)$.

Instead, a probabilistic greedy algorithm would aim to reduce the number of combinations by randomly selecting initiators. That is, at each step $t$, in order to measure the impact of a node $i$ in the presence of other nodes, $i$ would have to be selected as an initiator. Then, the remaining initiators would be randomly selected. This process would be repeated $v$ times, each time recording the cascade size. We would have essentially measured the impact of node $i$ as an initiator in the presence of randomly selected set of initiators. We would repeat this process for all other inactive nodes, and finally select the node with the highest impact.
Since we would have to measure the impact of each inactive node, and run $v$ simulations to do so, the time complexity per step $t$ is $\mathcal{O}\left(vN\right)$. Typically, $g$ is proportional to the total number of nodes $N$, adding an additional  $\mathcal{O}\left(N\right)$. Finally, computing the simulations for each set of initiators takes another $\mathcal{O}\left( \langle k\rangle N\right)$, and so the total complexity of the probabilistic greedy algorithm for $g$ initiators is $\mathcal{O}\left( v\langle k\rangle N^3\right)$, which is still very expensive.

Here, we introduce the Group Performance Index algorithm (GPI) Alg.~\ref{Alg}. With GPI we target the nodes which, when included in any randomly selected initiator set (group), cause the group to have higher than average performance.
GPI shares similarities with the probabilistic greedy, however it is much more efficient. First, we take advantage of the property that permutations of any set of initiators do not impact the total cascade size returned by that set for the LTM. By not having to scan each node individually when computing its impact in the presence of other initiators, we reduce the number of computations by $\mathcal{O}\left( N\right)$. Moreover, for the probabilistic greedy algorithm we would be selecting each initiator one-by-one, each time having to update the impact of each node by re-running simulations. 

Instead, we select $\ceil{sN}$ initiators (instead of one-by-one, where $s\in(0,1)$) defines granularity of initiator set increases, thus reducing the complexity by another $\mathcal{O}\left( N\right)$ factor. Also, when randomly picking nodes as test-initiators for our metric, we do not predefine the order of selecting initiators, but we pick them randomly one by one from the inactive nodes.
Finally, we aim to maximize the cascade size for a specific number of initiators, which is essentially our cost budget. However, the impact of even a small fraction of initiators can potentially have a large impact on the cascade size, especially near a tipping point. That means, a nearby tipping point can be missed because of fixed seed budget.
To address this, we aim to minimize the size of the initiator set in order for the cascade size to be at least as large as a specific predefined fraction of nodes, $S_{\rm goal}$. However, in general, GPI can also be used when constraining either the seed budget, or computational time.

Let us start with the initial graph $G(V,E,r)$, where $V(G)$ is the set of $N$ nodes, $E(G)$ is the set of edges of the graph, while $r_i$ is the resistance of each node $i\in V(G)$. 
Our goal is to find the initiator set $Y$ such that the fraction of nodes activated by it satisfies the inequality $f(Y)/N\ge S_{\rm goal}$, where the function $f(X)$ expresses the number of nodes that got activated from the set $X$, see [Alg.~\ref{Alg}]. To do so, at each step $t$ we select the set $Q$ of $\ceil{sN}$ nodes with the highest GPI-ranking (we will define it below) as initiators. Then, we compute the cascade induced by $Q$ on the graph with cascade induced by nodes in $Y$ already included in graph $G$ and we include the $Q$ set of nodes in the initiator set $Y$ by updating $Y=Y\cup Q$. Finally, we update the spread size, and update $G(V,E,r)$ (reducing the resistance, and removing all activated nodes and their edges). 

As stated above, at any step $t$ we look for the nodes with the highest GPI value, that is, the nodes which when present in the initiator set cause the desired cascade size to be reached (on average) fast so with small initiator set size. To measure the expected GPI, a large enough number $v$ of simulations is required. Each simulation $j$ is run on a series of graphs with initial one defined as $G_{\rm test}=G$. This simulation stops when the desired cascade size  $\ceil{S_{\rm goal}N}$ is reached. For each simulation $j$, we randomly select one-by-one an inactive node, place it on the initially empty set $Y_{{\rm test},j}$, then we compute the spread, and update $G_{\rm test}$ accordingly. Each simulation stops when $f(Y_{{\rm test},j})\ge S_{\rm goal}N-f(Y)$.

Let $n_{j,i}$ be $1$ when $i\in Y_{{\rm test},j}$, and $0$ otherwise, informing whether node $i$ belongs to the set $Y_{{\rm test},j}$. At the beginning of each simulation, $Y_{{\rm test},j}$ is empty, hence, $n_{j,i}$ for each node $i$ is initially zero. Then, for each node $i$ that belong to
$Y_{{\rm test},j}$ the GPI metric can be expressed analytically as
\begin{equation}
\label{Eq2}
{\rm GPI}_i=\frac{\sum_{j=1}^v \left|Y_{{\rm test},j}\right|n_{j,i}}{\sum_{j=1}^v n_{j,i}},~{\rm for~} i {\rm~such~that}~\sum_{j=1}^v n_{j,i}>0
\end{equation}

The numerator of GPI$_i$ is the sum of sizes of the randomly selected initiator sets $\left|Y_{{\rm test},j}\right|$ in which node $i$ is present. Since we select inactive nodes uniformly at random, the nodes do not appear in the initiator sets equal number of times. If that was not the case (that is if all nodes were equally frequently chosen just like for the probabilistic greedy we mentioned above)  the numerator of the fraction of Eq.~\ref{Eq2} would be a sufficient metric, where the smaller the sum is the larger is the impact of node $i$. The presence of the denominator is necessary to normalize the number of times node $i$ is selected as an initiator. In addition, because we only select inactive nodes, the nodes which are more likely to be activated through spread, that is typically nodes with low resistance and high in-degree, are going to be less frequently a part of the initiator set  than others. And so, nodes with a large number of appearances are nodes less likely to be activated by diffusion than others.
 
 \begin{algorithm}
\caption{Group Performance Index Algorithm}
\label{Alg}
\begin{algorithmic}
\Procedure{Group Performance Index Algorithm}{}
\State{Input Graph $G\left(V,E,r\right)$ and thresholds $\phi_i$ for each node $i$}
\State{Input the desired cascade fraction $S_{{\rm goal}}$, step granularity fraction $s$, number of randomizations $v$}
\State{Initialize the initiator set $Y=\emptyset$} and cascade fraction $S_y=0$
\State{Get the resistance of each node, $r_i=\ceil{\phi_ik^{{\rm in}}_{i}}$}

\State{Initiate step counter $t=0$}
\While {$S_y<S_{{\rm goal}}$}
\State{$t \gets t + 1$}
\State{Initialize GPI$_i$ \Comment{average impact GPI for node $i$}}
\State{Initialize $nu_i$ \Comment{numerator of Eq.~\ref{Eq2} for node $i$}} 
\State{Initialize $de_i$ \Comment{denominator of Eq.~\ref{Eq2} for node $i$}}
\State{Initialize $Q=\emptyset$ }
\For{$j=1:v$}
\State{$G_{{\rm test}}=G$}
\State{Initialize $Y_{\rm test}=\emptyset$ \Comment{set of test-initiators}} 
\State{Initiate the local test cascade fraction $S_l=0$}

\While {$S_l<S_{goal}-S_y$}
\State{Randomly select an inactive node $i$ as test initiator}
\State{$Y_{\rm test} \gets Y_{\rm test}\cup \{i\}$ \Comment{Add node $i$ to the initiator set $Y_{\rm test}$}}
\State{Run on $G_{{\rm test}}$ the cascade induced by $i$:}
\State{Compute the additional cascade size $f(i)$, and update $S_l$, $S_l \gets S_l + f(i)/N$}
\State{Add all newly activated nodes to $Y_{\rm test}$}
\State{$de_{i\in Y_{\rm test}} \gets de_{i\in Y_{\rm test}}+1$, $nu_{i\in Y_{\rm test}} \gets nu_{i\in Y_{\rm test}}+\left| Y_{\rm test}\right|$}
\State{Update graph $G_{\rm test}$ (update node resistance, remove the actived nodes and their edges)}
\EndWhile
\EndFor

\State{GPI$_i=nu_i/de_i$, for all $i \in Y_{\rm test}$}
\State{Insert the top $\ceil{sN}$ GPI-ranking nodes to $Q$, and $Y$, updating $Y \gets Y \cup Q$}
\State{Run the cascade on $G$ induced by the $Q$:}
\State{Compute the additional cascade size $f(Q)$ and update $S_y \gets S_y + f(Q)/N$}
\State{Update graph $G$ (update node resistance, remove activated nodes and their edges)}
\EndWhile
 \EndProcedure
\end{algorithmic}
\end{algorithm}

Since GPI deals with the expected impact of nodes, it is by default slower than the rest of  the strategies but can potentially find much better initiator sets than other strategies do. 
Computing the GPI of each node takes $\mathcal{O}\left(v\langle k\rangle N^2\right)$, since, we randomly select one-by-one test-initiators $\mathcal{O}\left(N\right)$, compute the spread $\mathcal{O}\left(\langle k\rangle N\right)$ (which for sparse graphs becomes $\mathcal{O}\left(N\right)$) until the $S_{\rm goal}$ is satisfied, and finally we repeat the process $v$ times. Sorting the nodes based on their GPI ranking takes $\mathcal{O}\left(\log N\right)$ using a heap structure. Finally, for each $Q$, for each of the  $\mathcal{O}\left( N\right)$ highest GPI-ranked nodes we compute the spread size in $\mathcal{O}\left\langle k\rangle N\right)$. 
Thus, the complexity for each batch of seeds $Q$ is $\mathcal{O}\left(v\langle k\rangle N^2 + \log N + \langle k\rangle N^2\right)$, so $\mathcal{O}\left(v\langle k\rangle N^2\right)$. 
Since the final size of the seed set $|Y|$ is $\mathcal{O}\left(\log N\right)$, we require $\lceil |Y|/\lceil sN\rceil\rceil$ steps which is $\mathcal{O}\left(1/s\right)$, hence the final complexity is $v/s*\mathcal{O}\left(\langle k\rangle N^2\right)$ which reduces to $\langle k\rangle v/s\mathcal{O}\left(N^2\right)$ for sparse graph for which $\langle k\rangle$ is a constant not changing with $N$.


From our experience, for a good estimation of the expected GPI value of each node the number of randomizations $v$ should be of the order of the system size. In addition, it's possible to control the size of the computation of the spread any initiator would induce on a sphere of influence $L$. Then, under the assumption of sparse graphs the spread computation is an $\mathcal{O}\left( 1\right)$ process, hence the total complexity becomes 
$v/s\langle k\rangle\mathcal{O}\left(N\right)$, becoming competitive in terms of complexity with the best known solutions. Furthermore, for the case of an identical threshold for all nodes, since there is a sharp phase transition point (and small spread otherwise), the computation of spread size is even faster. For the reminder of the paper, unless otherwise specified, we set the control parameters of GPI as $s=10^{-3}$, $v=10^5$, and $S_{\rm goal}=0.5$, and will compute the full spread size induced by any initiator node. Those value were chosen based on performance of alternatives. Indeed, $v/s$ defines the computational budget for finding a solution, So for fair comparison of different values of $v$ and $s$, we need to keep $v/s$ constant. Hence, a smaller value of $s$ means that $v$ needs to be kept low resulting in poor selection of seeds due to unreliable estimate of GPI. In contrast, large $s$ means that we have large $v$ which results in good selection of seeds, but also large $Q$, so these seeds are likely to become obsolete before they are replaced because graph may change much after less then $|Q|$ seeds are added. Finally, when $s>1/N$, then there is a non-zero probability that a (small fraction) of selected initiators will be activated from a previously selected initiator. For those cases, we excluded the activated initiators from the final initiator set.

\section*{Results}
We are comparing the performance (cascade size $S_{\rm eq}$) of the strategies for the entire parameter space of network assortativity $\rho$, and threshold distribution with fixed average threshold $\overline{\phi}=0.50$ and varying standard deviation $\sigma$. In Fig.~\ref{Fig1} we present our main results for the ensemble of ER graphs for the extreme cases of high positive ($\rho=0.9$) and high negative ($\rho=-0.9$), and neutral ($\rho=0$) assortativity, measured with Spearman’s rank correlation coefficient $\rho$~\cite{Hovstad}. Furthermore, we examine the cases of a identical thresholds ($\sigma=0$), a uniform threshold distribution ($\sigma=0.287$), and a truncated normal distribution in between ($\sigma=0.2$). For the GPI strategy we present the critical initiator fraction for which the $S_{\rm goal}$ is satisfied. First, focusing especially on ER graphs ($\rho=0$) we notice that as we move from a threshold distribution with standard deviation $\sigma=0$ to larger $\sigma$, there is a change from a first-order phase transition to a smooth crossover also seen for randomly selected initiators in~\cite{Karamp_PLOSONE}. Interestingly, in the case of the uniform threshold distribution ($\sigma=0.287$) at the ensemble level, we observe that all the direct methods appear to have diminishing returns with increasing cascade size, that is, the contribution to the cascade size of any additional initiator in an initiator set is diminishing as the initiator set is becoming larger.

As far as the performance of the strategies is concerned, the degree ($deg$) strategy's relative performance is decreasing for larger $\sigma$'s, while the resistance strategy's performance is increasing. In addition, CI-TM which incorporates a network structure decomposition using the neighboring nodes with the information about resistance $v=1$ in it as metric, is out-performing the degree strategy for the case of $\sigma=0$ and $\rho=0$ but it is not performing as well in the rest of cases. On the other hand, the ID approach is outperforming the degree, the resistance and the CI-TM strategy for all cases of $\rho$ and $\sigma$. 
Naturally, the introduced weighted strategy is outperforming in all cases the strategies that incorporates their ranking metrics ($deg$, $res$, CI-TM, DD, ID), especially for $S_{eq}=0.5$ which is what we optimized it for here.

The GPI strategy largely outperforms the rest of the strategies in all cases except for the case of very high assortativity $\rho$ with identical thresholds ($\sigma=0$). Yet, for a desired cascade size of $S_{eq}=0.5$ with even lower $s$ and higher $v$ the performance of GPI improves. The strong performance of GPI in the case of ER graphs ($\rho=0$) with a identical threshold $\sigma=0$ reveals the importance of the combination of initiators for Influence Maximization. We have used a $s=10^{-3}$ fraction of initiator and $v=10^5$, even though the best resolution we could achieve for a graph with size $N$ would be $s=1/N$, because then the initiators are inserted one-by-one.

To further study the direct methods, we evaluate their performance in the assortativity space $\rho$ (Fig.~\ref{Fig2}) and the threshold distribution space $\sigma$ (Fig.~\ref{Fig3}) for $S_{eq}=0.5$. Evaluating through $\rho$ for $\sigma=0$, we observe a highly non-monotonic behavior for all the strategies. Between the range of approximately $-0.8\leq \rho \leq0.4$, CI-TM is outperforming the degree strategy, which is approximately in the same regime, in which ID is outperforming all other direct strategies. Yet, for $\rho\ge0.5$ DD is the best strategy. With increased deviation of the thresholds $\sigma=0.20$ (Fig.~\ref{Fig2}b), the performance of the strategies which depend more on the network structure like the degree and CI-TM is getting worse, while strategies which give higher importance to the resistance of the node, like $res$, DD and ID are performing better. Finally, for a uniform threshold distribution $\sigma=0.2887$ (Fig.~\ref{Fig2}c), we observe that strategies show a convex response to $\rho$, with ID being the best strategy (for a desired cascade size $S_{{\rm goal}}=0.5$). On the other hand, the threshold ($thres$) strategy appears to be independent of $\rho$ for thresholds generated randomly with $\sigma=0.2887$. Finally, from Fig.~\ref{Fig3} we observe that ID has the best performance compared to all the direct strategies over nearly all of the entire range of $\sigma$, making it the best overall strategy (excluding the weighted and GPI strategies). 

In order to better study the performance of each strategy, we additionally examined the probability of any strategy being the best one (Fig.~\ref{Fig4}) for the same initiator fraction $p$. In particular, we tested the performance of all strategies at a fixed randomly generated ER network, for 200 different threshold generations. The actual cascade size $S_{eq}$ vs. initiator fraction $p$ can be seen in the Supplementary Material Fig.~\ref{FigS1}a (average) and Fig.~\ref{FigS1}b (50 runs). GPI remains superior even for a very small number of initiators.

First, we notice that the greedy algorithm is in all cases outperforming all other strategies for a small initiator fraction, while for a very large initiator fraction all strategies have the same performance, because all of the network is activated. For an ER graph ($\rho=0$) with $\sigma=0$ the spread is minimal until the phase transition point has been reached, and CI-TM leads until the tipping point for ID has been reached. Since the metric of CI-TM takes into account the out-degree of the nodes, and does not consider the effort required to activate nodes with high resistance and low in-degree, it is outperforming other methods for smaller initiator sizes, but eventually gets surpassed, when the nodes with high resistance with a structural importance have not been activated.
However, for very low ($\rho=-0.9$) and high  ($\rho=0.9$) degree assortativity, there are multiple initiator fractions for which a large spread occurs, which vary for the strategies, allowing for a sudden change of `lead' between the four network depend strategies (CI-TM, ID, $deg$, and DD), with DD and ID performing the best. For larger $\sigma$ the importance of resistance increases while the importance of the network structure declines, making the strategies more depended on resistance to take the `lead', while also reducing the number of sharp `lead' changes.

Next we focus on the BI and GPI strategies and their performance for their different parameters. For the BI strategy, we scan the parameters space $a \times b$ and record the average minimum $p_c$ at which the cascade size is $S_{{\rm goal}}=0.5$ (Fig.~\ref{Fig5}). The second and third feature of this weighted method defined by Eq.~\ref{Eq1} correspond to the first two terms of the CI-TM metric. The three features are interdependent, e.g. before the cascade begins with $\sigma=0$, a node's $i$ resistance $r_i$ is nearly linearly proportional to its degree $k_i$, which is why we observe those linear contours on the plots a, d, e. Moreover, it is clear that any strategy that would exclude the resistance ($a=0$) from its metric, such as the CI-TM, will have inferior performance. The contours indicate that the importance of the resistance and degree of a node relative to the third feature, which in addition is computationally most costly to obtain. Furthermore, we have recorded the impact of the standard deviation $\sigma$ on the optimal weights (Fig.~\ref{Fig6}). As $\sigma$ increases, the optimal $c$ coefficient decreases. Interestingly, the most important feature is the resistance ($a\approx0.53$), then the degree ($b\approx0.32$), and the smallest importance is left for  $c\approx0.15$. This result is especially important since other strategies do not fully utilize the resistance information combined with other network centrality measures.

For the GPI strategy, Fig.~\ref{Fig7} and Fig.~\ref{Fig8} we show its performance of GPI in relationship to the number of randomizations $v$ and step sizes $s$, respectively.
On average, with increasing $v$ and decreasing $s$ we always minimize $p_c$ for obtaining a cascade desired size (here we aim at the size $N/2$). On average, we expect and observe an asymptotic return with increasing $v$. For computational efficiency, we fix the $s$ and $v$ when the additional performance is minimal. Further investigation is required in order to find the interplay between the two control parameters in order to optimize the performance of the algorithm for the smallest computational time possible. Interestingly, for $\sigma=0.2878$ in contrast to the direct methods, there is a large transition on cascade size as we reach $S_{{\rm goal}}$.

\section*{Discussion}


The challenge of Influence Maximization for the LTM or other diffusion processes is finding low complexity, yet well performing algorithms for the discovery of superspreaders. We tackle this challenge by introducing two different approaches. 
On our first approach we take advantage of the combination of different node features: the resistance to influence, the out- degree, and the impact of its activation to it's second neighborhood. Those features essentially capture the resistance drop of the network. By placing weights between them and constructing the BI metric, we are able to study the importance of each of those features for a variety of network structures (degree assortativity) and threshold distributions.
We discovered that resistance is the most impactful of the features. Hence, the cascade size is governed not by initiators with high network centrality measures but by low resistance nodes, a result also supported by Watts and Dotts \cite{Watts_2007}. 
In addition, we optimized those weights for influence maximization. Among the strategies with similar or lower complexity, BI always outperforms them for any degree assortativity or threshold distribution. Even in the case of non-optimized weights, like considering equal weights between the three features (ID strategy) or just the resistance and out-degree (DD), our method is out-performing all other strategies with similar or lower complexity ($\mathcal{O}\left(\langle k\rangle N\log N\right)$, which is reduced by a factor $\mathcal{O}\left(\langle k\rangle\right)$ for sparse graphs).



Yet, most strategies (including ours above), are based on heuristic/analytical metrics that do not take into consideration the combination of nodes, leading to local solutions of reduced size of initiators' set. 
Our second approach (GPI) focuses on the random combination of nodes for influence maximization thus targeting a global optimum. It allows us to control a number of parameters which in return control the performance and computational time. The time complexity for sparse graphs is $\mathcal{O}\left( vN^2\right)$ (which can be reduced by a factor $\mathcal{O}\left(N\right)$ using a limited sphere of influence for computing an approximation of the spread). Our results show that the GPI metric performs better than any strategy against we compared it for almost any initiator size, threshold distribution, and network assortativity. To say the least, GPI serves as a benchmark for (synthetic) graphs and sets a minimum bound for the optimal initiator set. Finally, in terms of applicability, both our approaches can be used for directed and weighted graphs as well, although a few adjustments would have to be made for the weighted strategy in case of weighted graphs.

Our two approaches, combining node features, and combining nodes, show there are more possible improvements to be made on both the performance and time-complexity for Influence Maximization. New methods could simultaneously take into account the network and model specific node properties as well as the combination of nodes. Other methods could possibly focus on learning to discover important node features for instance. As far as GPI is concerned, a number of improvements can be made for controlling the number of repetitions $v$ while keeping the algorithm performance approximately the same.

\section*{Methods}
For the generation of Erd\H{o}s-R\'enyi (ER) graphs~\cite{Erdos_1959} we used the $G(N,p_{\rm ER})$ model with $N$ being the system size and $p_{ER}$ the probability that a random node will be connected to any node in the graph. The probability $p_{ER}$ is given by $p_{ER}=\langle k\rangle/\left(N-1\right)$, where $\langle k\rangle$ is the nominal average degree in the network. The probability of the existence of a disconnected component is $~4.5\times10^{-5}$ for $\langle k\rangle=10$ and $N=10^4$. \\
The degree assortativity was first introduced by Newman~\cite{Newman_assortativity} to describe the connectivity between neighboring nodes with different degrees. To measure it we use Spearman's $\rho$~\cite{Hovstad}. ER graphs have $\rho=0$ degree assortativity. To control the degree assortativity, we use the method applied in~\cite{Molnar_SciRep2014}. 

As far as the thresholds are concerned, they are bound between 0 and 1. 
The threshold distributions are truncated to generate numbers between the above two bounds~\cite{Karamp_PLOSONE}.
The standard deviation for a uniform threshold distribution for these bounds is $\sigma=1/\sqrt{12}\approx0.2887$.
The truncated threshold distribution $P(\phi,\sigma)$ is given by
$P(\phi,\sigma)=N(\mu,\tau)/(1- \int_{-\infty}^{0}N(\mu,\tau) d\mu -\int_{1}^{\infty}N(\mu,\tau) d\mu)$ 
for $0 \le \phi \le 1$, and $P(\phi,\sigma)=0$ anywhere else, similar to~\cite{Karamp_PLOSONE}. 
In the above, $N(\mu,\tau)$ is the normal distribution with mean $\mu$ and standard deviation $\tau$, which take values $0 \le \mu \le 1$ and $0\le \sigma \le \infty$ respectively.

\section*{Acknowledgments}
The authors thank Prof. Radoslaw Michalski for his comments on this work.
This work was supported in part
by the Army Research Office grant W911NF-12-1-0546,
by the Army Research Laboratory under Cooperative Agreement Number W911NF-09-2-0053, 
by the Office of Naval Research Grant No.~N00014-09-1-0607 and N00014-15-1-2640,
and the National Science Centre, Poland, project no. 2016/21/B/ST6/01463. 
The views and conclusions contained in this document are those of the authors and should not be interpreted
as representing the official policies either expressed or implied of the Army Research
Laboratory or the U.S. Government.

\section*{Author Contributions}
P.D.K., B.K.S. and G.K. designed the research;
P.D.K. implemented and performed numerical experiments and simulations;
P.D.K., B.K.S. and G.K. analyzed data and discussed results;
P.D.K., B.K.S. and G.K. wrote, reviewed, and revised the manuscript.

\section*{Additional Information}
Competing financial interests: The authors declare no competing financial interests.

\section*{Supplementary Information}
\subsection*{Additional Info for probability area plot (Figure 4)}
We provide additional information for Fig.~\ref{Fig4}. We present the corresponding average performance and 50 first runs of each strategy for one ER network for different threshold generations in Fig.\ref{FigS1}a) and Fig.\ref{FigS1}b) respectively.
As is customary, all strategies are compared for the same threshold realizations. These results further support the increasing performance of GPI with increasing initiator fractions.

\newpage

\section*{Figures \& Captions}
\nopagebreak
\begin{figure}[tbh]
\centerline{\includegraphics[width=\textwidth]{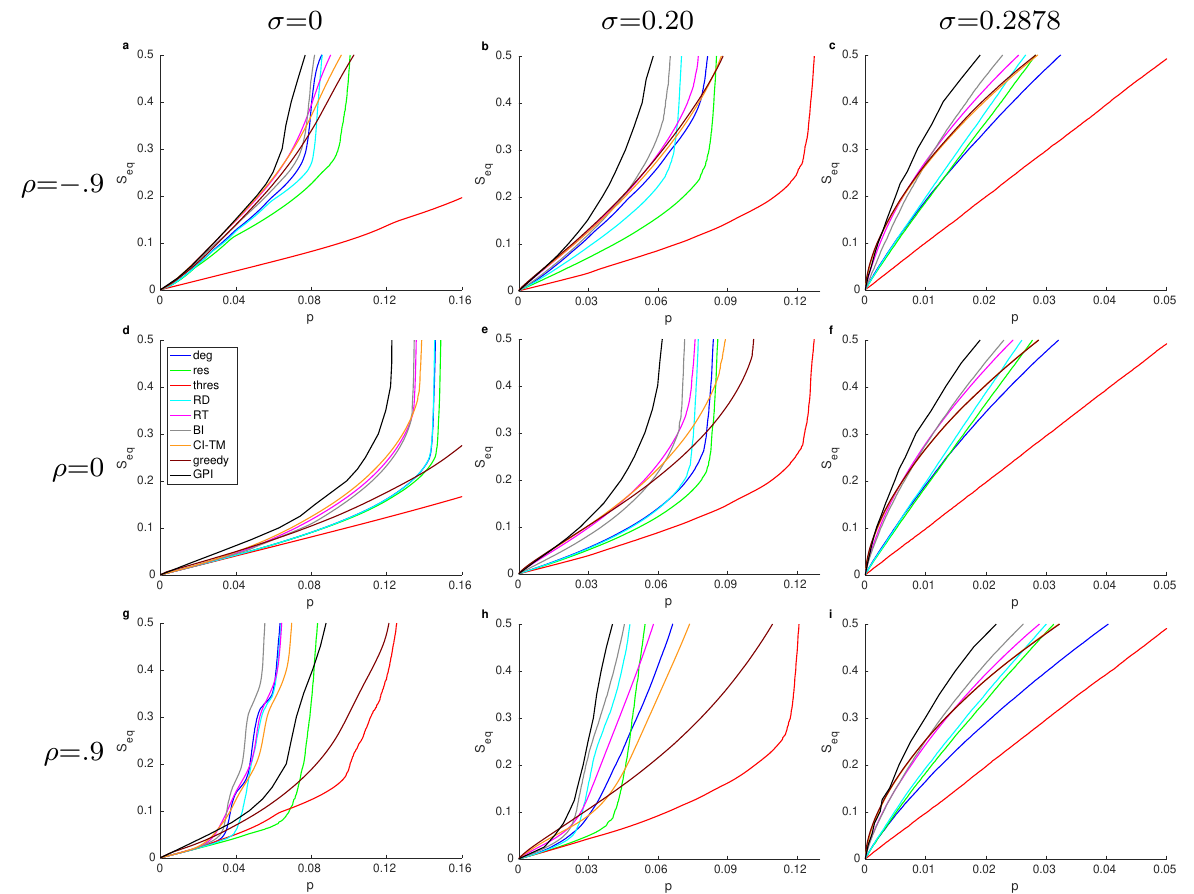}}
\caption{Comparison of the BI and GPI (for different $S_{{\rm goal}}$ selection strategies in terms of cascade performance $S_{eq}$ 
for \textbf{(a-b-c)} $\rho=-0.9$, for \textbf{(d-e-f)} $\rho$$=$$0$, for \textbf{(g-h-i)} $\rho$$=$$0.9$,
for \textbf{(a-d-g)} $\sigma$$=$$0$, for \textbf{(b-e-h)} $\sigma$$=$$0.20$, for \textbf{(c-f-i)} $\sigma$$=$$0.2887$,
with $\overline{\phi}$$=$$0.5$, averaged for 500 different network realizations (except for the GPI which is based on 20 realizations) each with a different threshold generation, applied on ER graphs with $N$$=$$10,000$ and $\langle k \rangle$$=$$10$.  For BI, the weights are optimized for $S_{\rm eq}$$=$$0.5$. For GPI, the fraction of top $s$ initiators, and number of randomizations are respectively $s$$=$$0.001$, and $v$$=$$100,000$,.}
\label{Fig1}
\end{figure}
\pagebreak

\begin{figure}[tbh]
\centerline{\includegraphics[width=\textwidth]{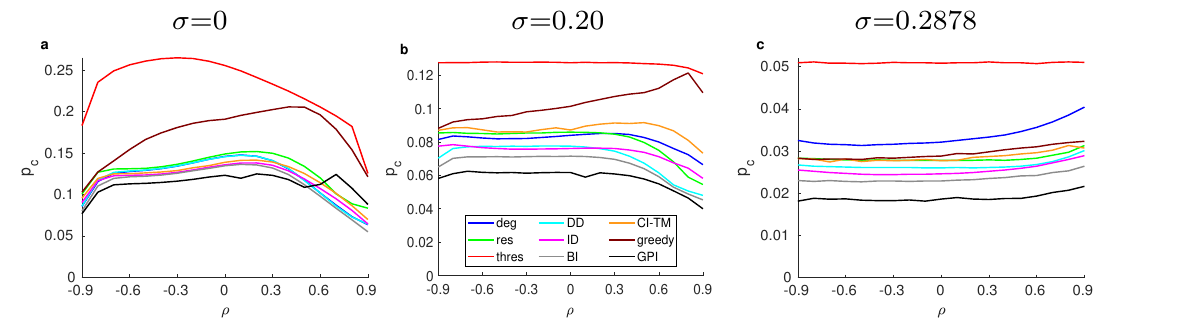}}
\caption{Initiator fraction $p_c$ required to reach spread $S_{eq}$$=$$0.5$ vs. degree assortativity $\rho$ for graphs with $N$$=$$10,000$ and $\langle k\rangle$$=$$10$ for 
(a) $\sigma$$=$$0$, for (b) $\sigma$$=$$0.20$, and for (c) $\sigma$$=$$0.2887$ with $\overline{\phi}$$=$$0.5$, averaged for 500 different network realizations (except for the GPI which is 20) each with a different threshold generation.}
\label{Fig2}
\end{figure}
\pagebreak

\nopagebreak
\begin{figure}[tbh]
\centerline{\includegraphics[width=\textwidth]{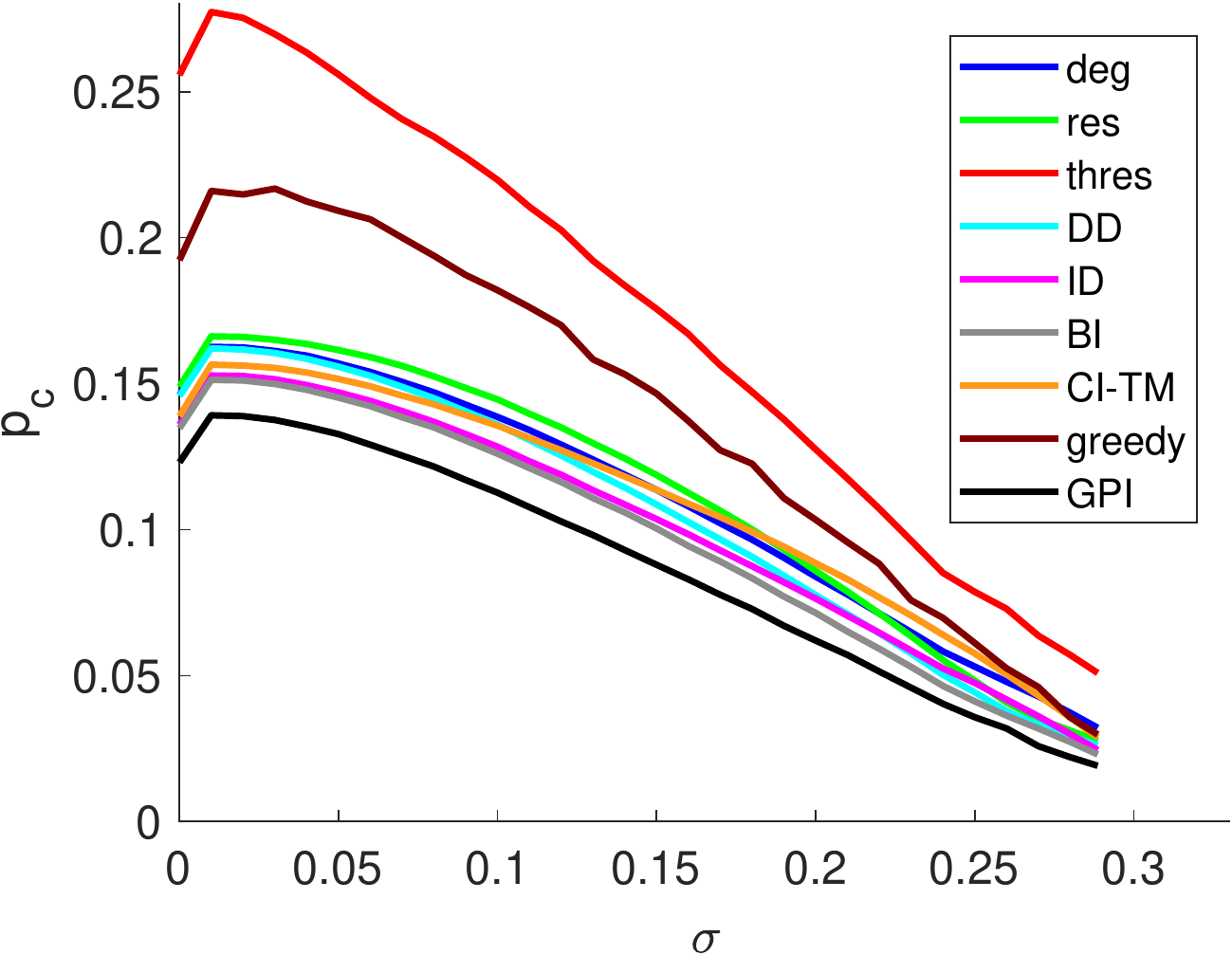}}
\caption{Initiator fraction $p_c$ required to reach spread $S_{eq}$$=$$0.5$ vs. the standard deviation $\sigma$ of the generated threshold distribution with $\overline{\phi}$$=$$0.5$ for graphs with $N$$=$$10,000$, $\langle k\rangle$$=$$10$ and
a $\rho$$=$$-0.9$, b $\rho$$=$$0$ c, $\rho$$=$$0.9$, averaged for 500 different network realizations (except for the GPI which based on 20 realizations) each with a different threshold generation.}
\label{Fig3}
\end{figure}
\pagebreak

\nopagebreak
\begin{figure}[tbh]
\centerline{\includegraphics[width=\textwidth]{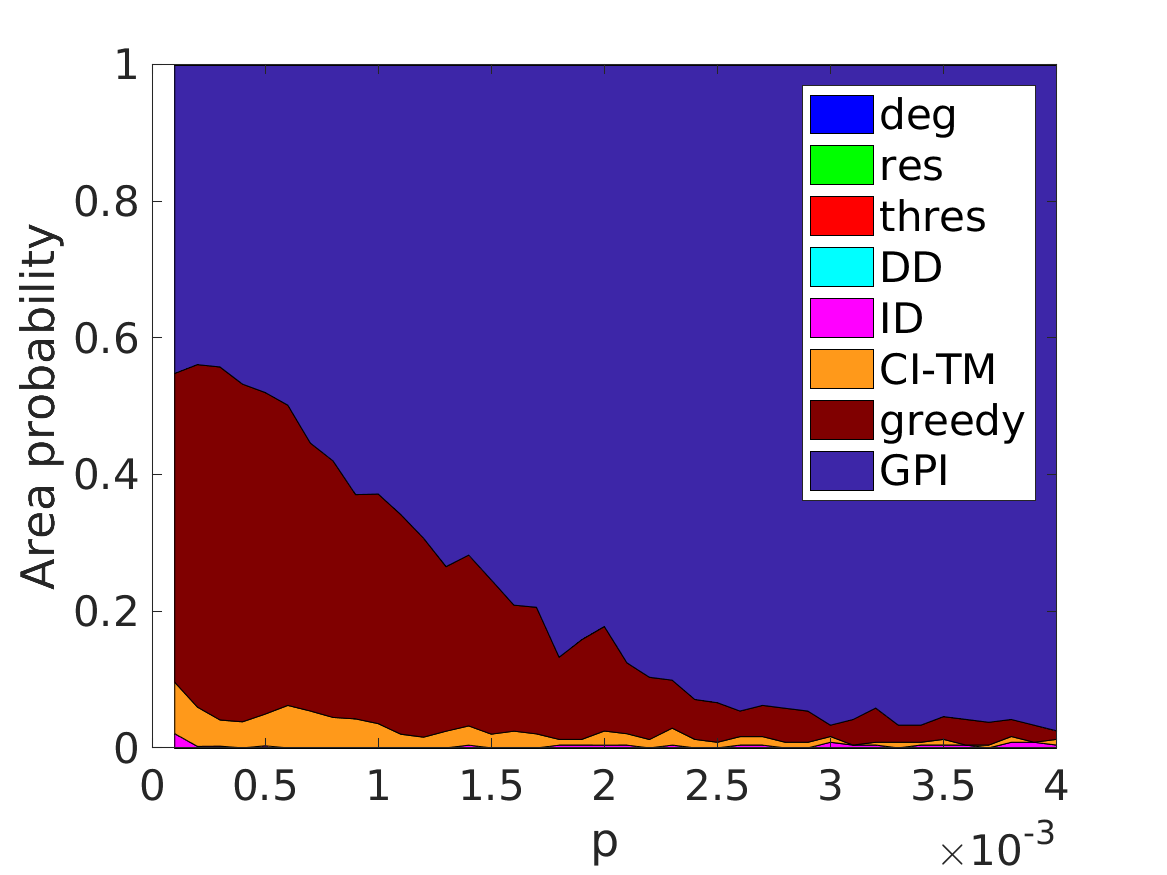}}
\caption{Probability of each strategy being the best strategy for one network with $N$$=$$10,000$, $\langle k\rangle$$=$$10$, $\rho$$=$$0$, $\sigma$$=$$0.2887$, and $\overline{\phi}$$=$$0.5$, for 240 threshold generations (same for each strategy).}
\label{Fig4}
\end{figure}
\pagebreak

\nopagebreak
\begin{figure}[tbh]
\centerline{\includegraphics[width=\textwidth]{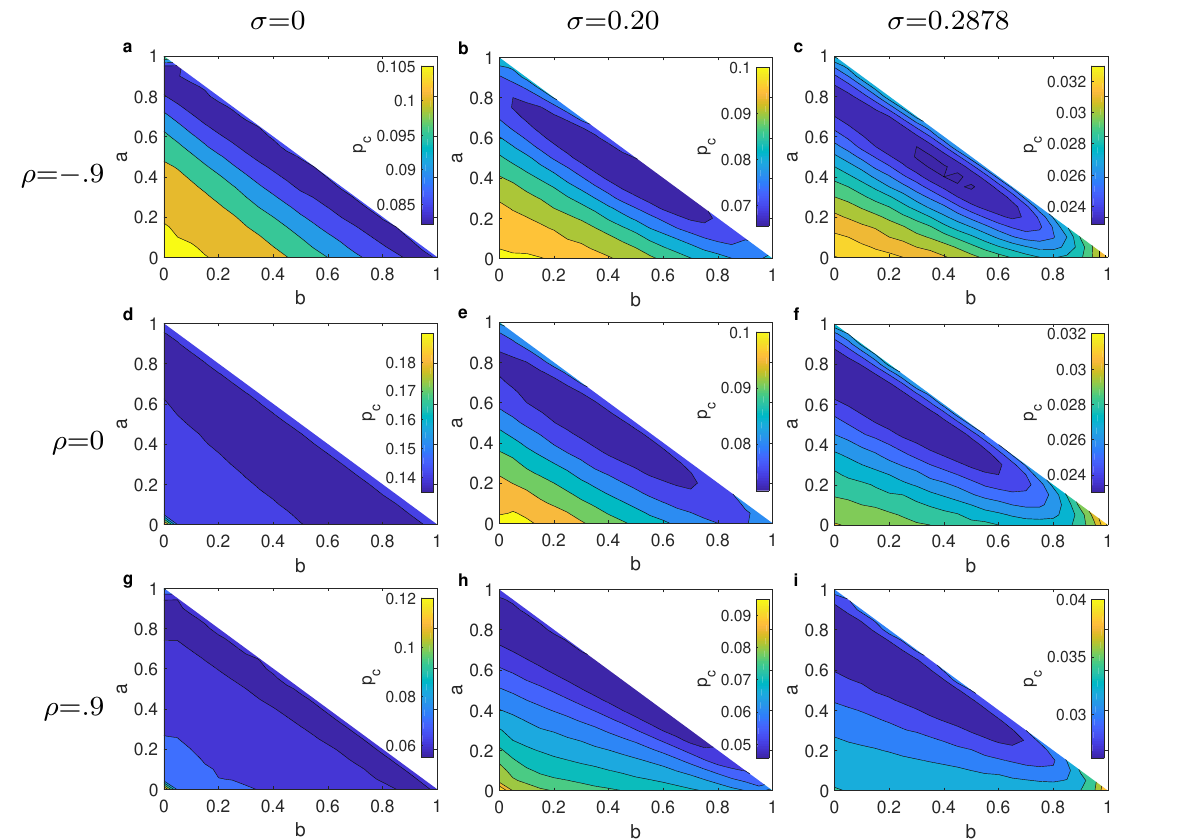}}
\caption{Contours of $p_c$ for reaching $S_{eq}$$=$$0.5$ by controlling parameters $a$ and $b$ (from Eq.~(\ref{Eq1}), with $a$$+$$b$$+$$c$$=$$1$) for graphs with $N$$=$$10,000$, $\langle k\rangle$$=$$10$ and  
for (a-b-c) $\rho$$=$$-0.9$, for (d-e-f) $\rho$$=$$0$, for (g-h-i) $\rho$$=$$0.9$, 
for (a-d-g) $\sigma$$=$$0$, for (b-e-h) $\sigma$$=$$0.20$, for (c-f-i) $\sigma$$=$$0.2887$,
with mean threshold $\overline{\phi}$$=$$0.5$, for 500 repetitions, averaged for 500 different network realizations each with a different threshold generation.
The resolution in the $a$ and $b$ weight space are 0.05.}
\label{Fig5}
\end{figure}
\pagebreak

\nopagebreak
\begin{figure}[h!]
\centerline{\includegraphics[width=\textwidth]{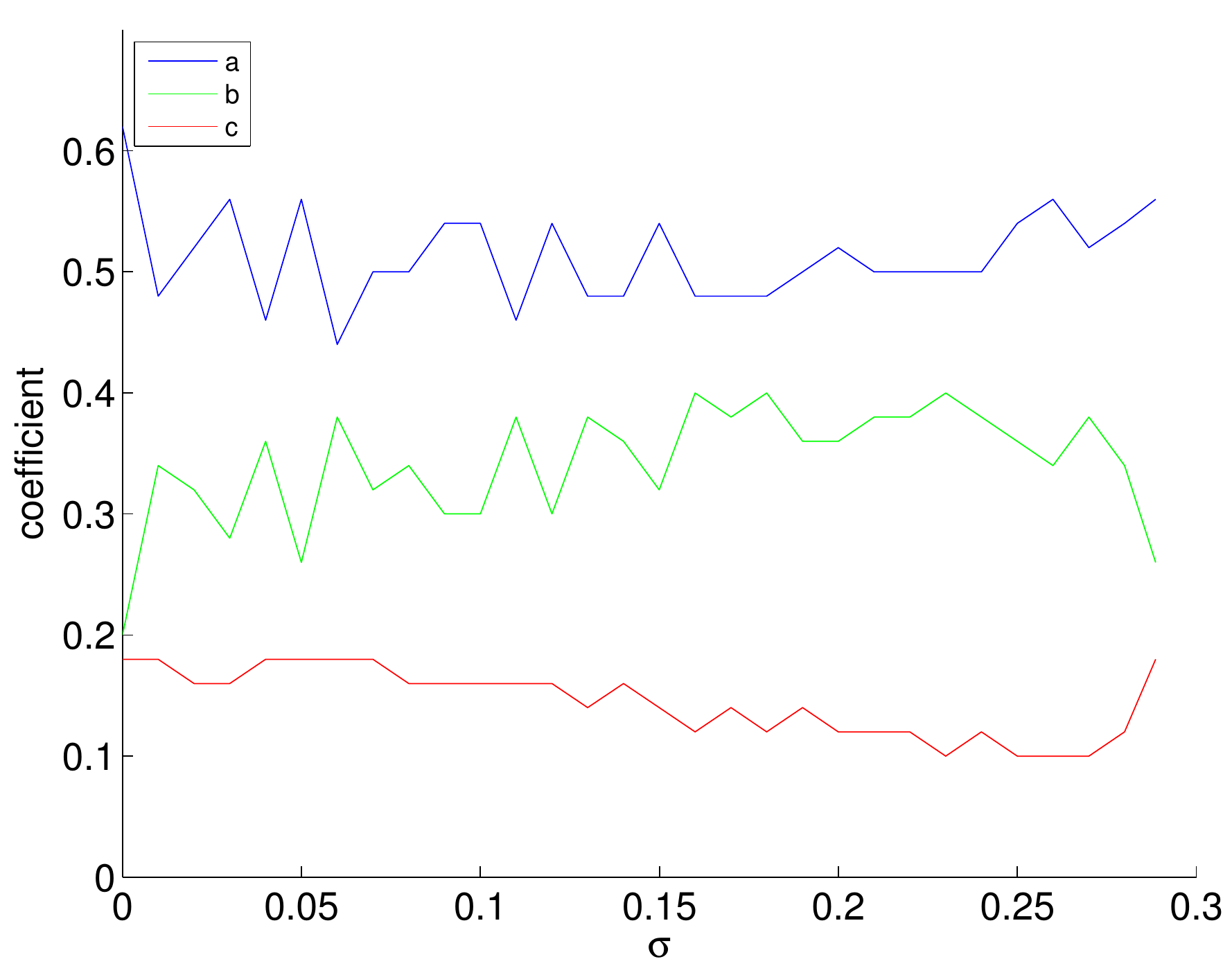}}
\caption{Impact of standard deviation $\sigma$ on the optimal weights (from Eq.~(\ref{Eq1}), with $a$$+$$b$$+$$c$$=$$1$)
for desired cascade $S_{{\rm goal}}$$=$$0.5$, for ER graphs with $N$$=$$10,000$, $\langle k\rangle$$=$$10$, $\rho$$=$$0$, with $\overline{\phi}$$=$$0.5$, averaged for 500 different network realizations each with a different threshold generation. The resolution in the $a$ and $b$ weight space are 0.02.}
\label{Fig6}
\end{figure}
\pagebreak

\nopagebreak
\begin{figure}[h!]
\centerline{\includegraphics[width=\textwidth]{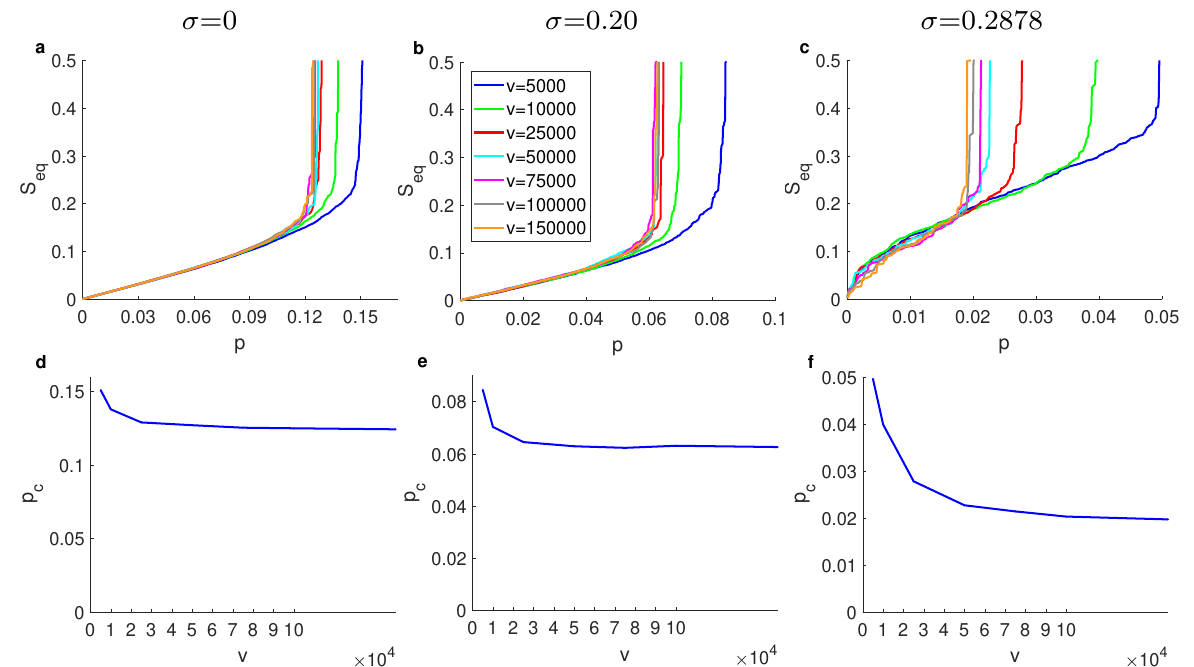}}
\caption{Impact of the number of randomizations $v$ on the performance of the GPI strategy for $s$$=$$0.0025$
for desired cascade $S_{{\rm goal}}$$=$$0.5$, for ER graphs with $N$$=$$10,000$, $\langle k\rangle$$=$$10$, $\rho$$=$$0$.
(a-b-c) cascade $S_{eq}$ vs. the initiator fraction $p$ for $\sigma=$$0$, $0.2$, $0.2887$ respectively (for one realization).
(e-f-g) initiator fraction $p_c$ required for desired cascade $S_{{\rm goal}}$$=$$0.5$ vs. randomizations $v$ for $\sigma$$=$$0$, $0.2$, $0.2887$ respectively (for one realization).}
\label{Fig7}
\end{figure}
\pagebreak

\nopagebreak
\begin{figure}[h!]
\centerline{\includegraphics[width=\textwidth]{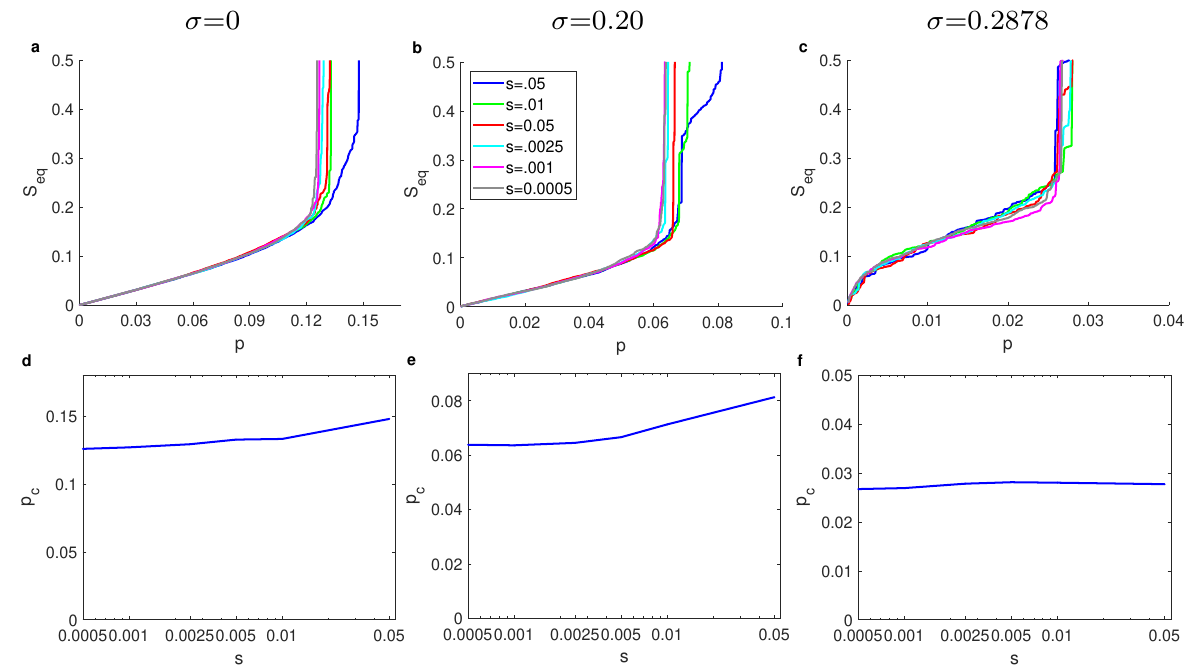}}
\caption{Impact of the size of initiator fraction $s$ on the performance of the GPI strategy for randomizations $v$$=$$25000$
for desired cascade $S_{{\rm goal}}$$=$$0.5$, for ER graphs with $N$$=$$10,000$, $\langle k\rangle$$=$$10$, $\rho$$=$$0$.
(a-b-c) cascade $S_{eq}$ vs.  the initiator fraction $p$ for $\sigma$$=$$0$, $0.2$, $0.2887$ respectively (for one realization).
(e-f-g) initiator fraction $s$ required for desired cascade $S_{{\rm goal}}$$=$$0.5$ vs. randomizations $v$ for $\sigma$$=$$0$, $0.2$, $0.2887$ respectively (for one realization).}
\label{Fig8}
\end{figure}
\pagebreak

\nopagebreak
\begin{figure}[tbh]
\centerline{\includegraphics[width=\textwidth]{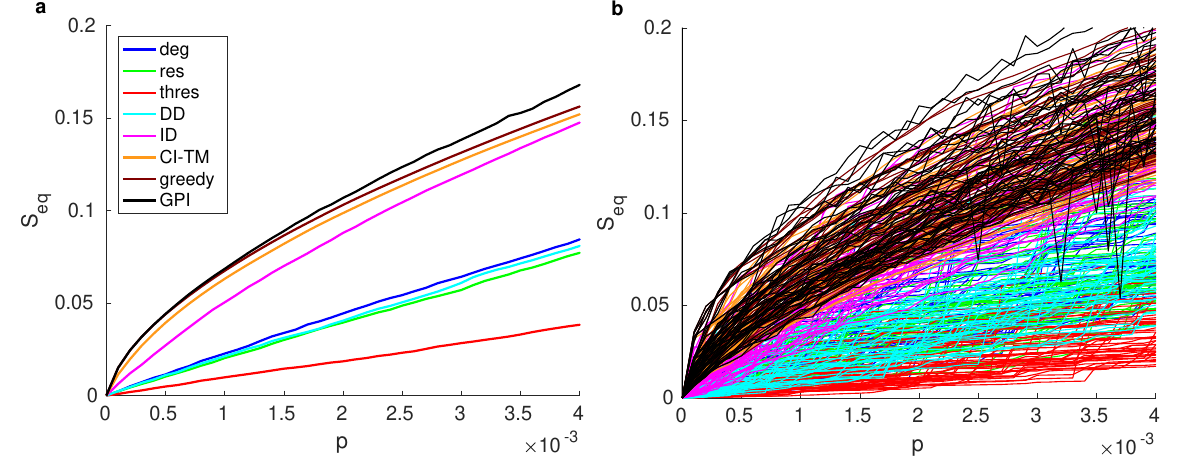}}
\caption{Comparison of cascade performance $S_{eq}$ for one network with $N$$=$$10,000$, $\langle k\rangle$$=$$10$ and
for (a) average performance for 240 threshold randomizations (same for each strategy), and 
for (b) first 50 threshold randomizations (same for each strategy),
for $\rho$$=$$0.9$, and $\sigma$$=$$0.2887$
with $\overline{\phi}$$=$$0.5$.}
\label{FigS1}
\end{figure}
\pagebreak
\end{document}